\documentclass[aps,pre,twocolumn,showpacs]{revtex4}
\usepackage{graphicx,epsfig}

\begin{document}
\title{Scaling of the Lyapunov exponent in type-III intermittent chaos}
\author{M. G. Cosenza} 
\affiliation{Centro de F\'isica Fundamental, Universidad de los
Andes, M\'erida, Apartado Postal 26 La Hechicera, M\'erida 5251, Venezuela.}
\author{O. Alvarez-Llamoza}
\affiliation{Centro de F\'isica Fundamental, Universidad de los
Andes, M\'erida, Apartado Postal 26 La Hechicera, M\'erida 5251, Venezuela.}
\affiliation{Departamento
de F\'isica, FACYT, Universidad de Carabobo, Valencia, Venezuela.}
\author{G. A. Ponce}
\affiliation{Departamento de F\'isica, Universidad Nacional Aut\'onoma de Honduras,
Tegucigalpa, Honduras.}


\begin{abstract}
The scaling behaviour of the Lyapunov exponent near the transition to
chaos via type-III intermittency is determined for a generic map. A
critical exponent $\beta$ expressing the scaling of the Lyapunov
exponent as a function of both, the reinjection probability and the
nonlinearity of the map is calculated. It is found that the critical
exponent varies on the interval $0 < \beta < 1$. This contrasts with
earlier predictions for the scaling behaviour of the Lyapunov
exponent in type-III intermittency.
\end{abstract}

\maketitle

\section*{1. Introduction}
Intermittency is a common scenario for the transition to chaos in
nonlinear dynamical systems. Intermittent chaos is characterised by
the display of long sequences of periodiclike behaviour, called the
laminar phases, interrupted by comparatively short chaotic
bursts. The phenomenon has been extensively studied since the
original work of Pomeau and Manneville \cite{Pomeau}
classifying type-I, -II, and -III instabilities when the Floquet
multipliers of the local Poincar\'e map associated to the system
crosses the unit circle. Type-I intermittency occurs by a tangent
bifurcation when the Floquet's multiplier for the Poincar\'e map
crosses the circle of unitary norm in the complex plane through
$+1$; type-II intermittency is due to a Hopf's bifurcation which
appears as two complex eigenvalues of the Floquet's matrix cross the
unitary circle off the real axis; and type-III intermittency is
associated to an inverse period doubling bifurcation whose Floquet's
multiplier is $-1$ \cite{Manneville}. Although other mechanisms may 
occur leading  to intermittency,  these three cases are the most simple 
and the most frequently encountered in low-dimensional systems.

Many experimental evidences for these types of intermittency have
appeared in the literature. In particular, type-III intermittency
has been found in lasers \cite{Tang}, electronic nonlinear
devices \cite{Testa,Fukushima,Ono}, biological tissues \cite{Griffith}, and
epilepsy \cite{Perez}. The statistical signature of intermittency
is usually given by the scaling relations describing the dependence
of the average length (denoted by $\left\langle l \right\rangle$) of
the laminar phases with a control parameter (denoted by $\epsilon$)
that measures the distance from the bifurcation point. Chaotic
dynamics is characterised by the positive sign of the largest
Lyapunov exponent (denoted by $\lambda$), although this quantity is
in general more difficult to measure from experimental data than
statistical variables such as the average laminar length. Pomeau and
Manneville \cite{Pomeau} assumed a uniform reinjection probability
into the laminar phase and calculated the scaling behaviour of both
the average laminar length and the Lyapunov exponent for those three
types of intermittent chaos. Specifically, for type-III
intermittency Pomeau and Manneville predicted the relations
$\left\langle l \right\rangle \propto \epsilon^{-1}$, and
$\lambda\propto \epsilon^{1/2}$, both when $\epsilon \rightarrow 0$.

However, theoretical and numerical studies \cite{Mayer1,Kawabe,Kodama1,Kodama2}
, as well as some recent experiments
\cite{Kahn,Ono,Griffith,Kim,Calvacante,Kye}, have shown deviations from the
Pomeau-Manneville's prediction for the scaling
behaviour of the average laminar length for type-III intermittency;
they found $\left\langle l \right\rangle \propto \epsilon^{-\nu}$,
with $1/2 < \nu < 1$.  These discrepancies are attributed to the
presence of mechanisms leading to nonuniform probability
reinjections. On the other hand, the scaling behaviour of the
Lyapunov exponent in type-III intermittency has rarely been
explored.

The aim of this paper is to clarify the effects of both the
reinjection probability and the nonlinearity on the scaling
properties of the Lyapunov exponent at the transition to chaos via
type-III intermittency by using a simple one-dimensional map. We
show that the behaviour of the Lyapunov exponent near this transition
exhibits appreciable deviations from the prediction of Pomeau and
Manneville. We find that $\lambda \propto \epsilon^{\beta}$, for
$\epsilon \rightarrow 0$, where the critical exponent $\beta$ varies
continuously in the interval $0 < \beta < 1$ as a function of the
map parameters.

\begin{figure}[ht]
\centerline{
\includegraphics[scale=0.95]{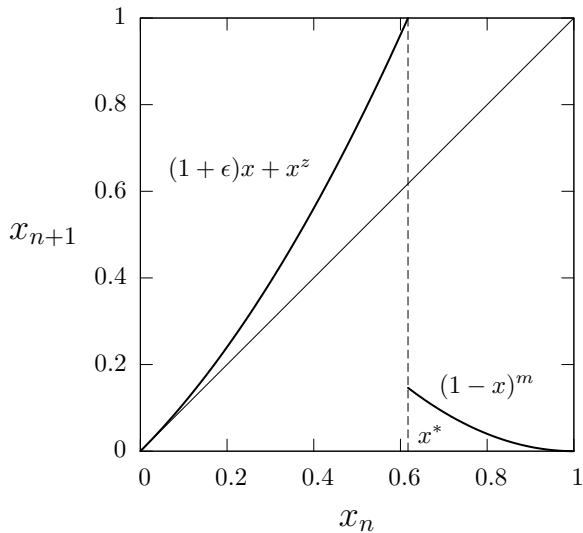}}
\caption{Generic map Eq.~(\ref{map}) for producing type-III
intermittency, with $z=2$, $m=2$ and $\epsilon=0.002$.}
\label{fig:1}
\end{figure}

\section*{2. Scaling Behaviour of the Lyapunov Exponent}
The return map is a fundamental information that can be obtained
experimentally in a chaotic system and inevitable contains some
reinjection process for type-III intermittency. Therefore, a
reinjection mechanism should be taken into account in the analysis
of characteristic quantities for intermittency such as the mean
laminar length or the Lyapunov exponent. We consider the following
one-dimensional map \cite{Kawabe}

\begin{equation}\label{map}
x_{n+1}= f(x_n)=\left\{
\begin{array}{lll}
 (1+\epsilon)x_n+x^z_n ,  & \mbox{if} & x_n < x^*\\
  (1-x_n)^m ,& \mbox{if} &   x_n > x^* \, ,
\end{array}
\right.
\end{equation}
where $x_n \in [0,1]$, $z \geq 2$, $m \geq 1$ and $x^*$ is the solution of
\begin{equation}
\label{x}
    (1+\epsilon)x^*+x^{*z}= 1 \, .
\end{equation}

The parameter $\epsilon$ measures the distance from the bifurcation
point. For $\epsilon > 0$, the map Eq.~(\ref{map}) generates
type-III intermittency near the origin which becomes an unstable
fixed point. The parameter $m$ controls the reinjection probability
into the laminar region of the map, with $m=1$ corresponding to
uniform reinjection. The reinjection probability becomes more
densely localised near the origin as $m$ increases. Figure \ref{fig:1}
shows the map Eq.~(\ref{map}).

The Lyapunov exponent is calculated as
\begin{equation}\label{Lyap}
\lambda=\lim_{N
\rightarrow\infty}\frac{1}{N}\sum_{n=0}^{N-1}\log|f'(x_n)|.
\end{equation}
We have used Eq.~(\ref{Lyap}) with $N=10^6$ for each set of
parameter values $z$, $m$ and $\epsilon$, averaging over $10^3$
random initial conditions $x_0$ uniformly distributed on the
interval $(0,1)$, after discarding $4 \times 10^3$ iterates as
transients. Figure~2 shows the Lyapunov exponent for the map
Eq.~(\ref{map}) as a function of the parameter $\epsilon$, for fixed
$z$ and several values of $m$, and for a fixed value of $m$ and
different values of the nonlinearity exponent $z$.

\begin{figure}[t]
\centerline{
\includegraphics[scale=0.95]{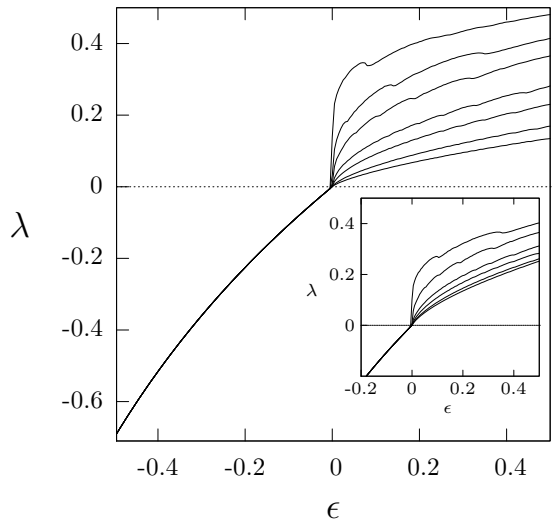}}
\caption{Lyapunov exponent $\lambda$ as a function of $\epsilon$ for
different values of $m$ and with fixed $z=2$; from top to bottom
$m=1,1.5,2,3,4,6$ and $8$. Insert: $\lambda$ as a function of
$\epsilon$, with fixed $m=2$, for different values of $z$; from top
to bottom $z=2,3,4,5,6$ and $8$.}
\label{fig:2}
\end{figure}

The onset of type-III intermittency is associated to a direct
transition from a fixed point to a chaotic one-band attractor. The
map Eq.~(\ref{map}) exhibits robust chaos for $\epsilon
> 0$. A chaotic attractor is said to be robust if, for its parameter
values, there exist a neighbourhood in the parameter space with
absence of periodic windows (i.e. with $\lambda > 0$) and the
chaotic attractor is unique \cite{Banerjee}. Robustness is an
important property in applications that require reliable operation
under chaos, in the sense that the chaotic behaviour cannot be
destroyed by arbitrarily small perturbations of the system
parameters. It should be noted that robust chaos not associated to
type-III intermittency has also been discovered in smooth,
continuous one-dimensional maps \cite{Andrecut}.

\begin{figure}[t]
\begin{center}
\includegraphics[scale=0.95]{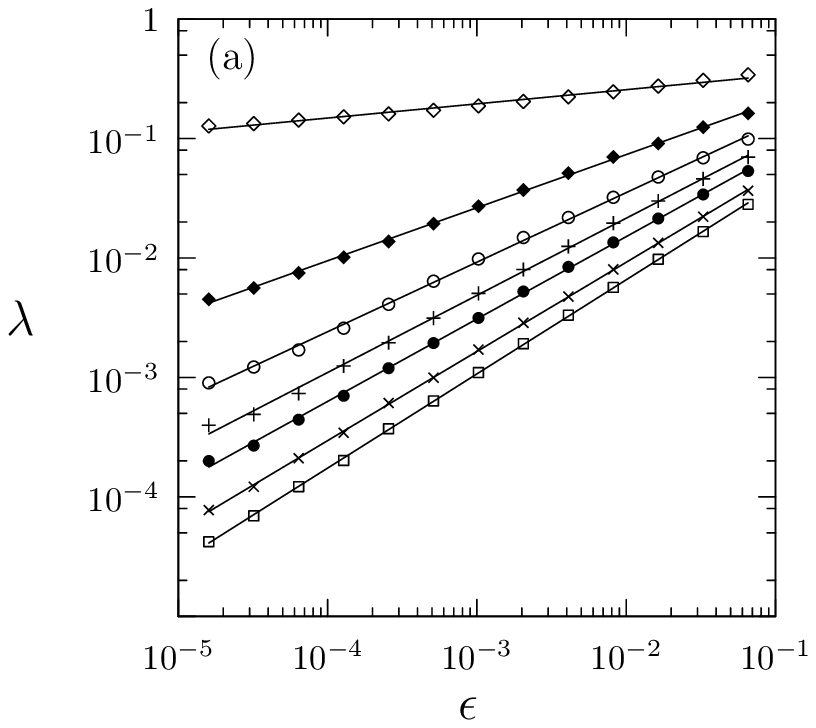}\\
\vspace{0.5cm}
\includegraphics[scale=0.95]{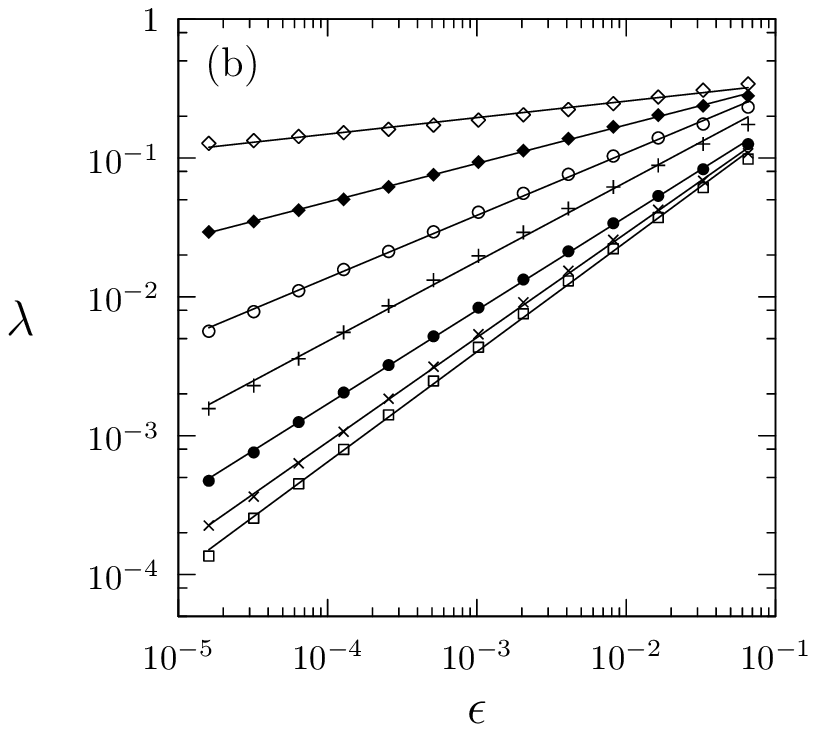}
\end{center}
\caption{(a) Log-log plot of the Lyapunov exponent $\lambda$ vs.
$\epsilon$ with fixed $z=2$ for different values of $m$; from top to
bottom $m=1,2,3,4,5,7$ and $9$. (b) Log-Log plot of $\lambda$ vs.
$\epsilon$ with fixed $m=1$ for different values of $z$; from top to
bottom $z=2,2.4,3,4,6,8$ and $10$.}
\label{fig:3}
\end{figure}

The transition to chaos via type-III intermittency is manifested by
a discontinuity of the derivative of the Lyapunov exponent at the
bifurcation value $\epsilon=0$. This discontinuity is due to the
sudden loss of stability of the fixed point associated to the
inverse period doubling bifurcation that occurs at the onset of
type-III intermittency. The Lyapunov exponent can be regarded as an
order parameter that characterises the transition to chaos via
type-III intermittency. This transition can be very abrupt as seen
in \ref{fig:2}. Thus, the behaviour of the Lyapunov exponent for $\epsilon
\rightarrow 0$ can be described by a scaling relation
\begin{equation}
\label{scale} 
\lambda \sim \epsilon^{\beta(z,m)},
\end{equation}
where $\beta$ is a critical exponent expressing the order of the
transition and that depends on the parameters $z$ and $m$.
Figure \ref{fig:3}(a) shows a log-log plot of the Lyapunov exponent vs.
$\epsilon$ for a fixed value of $z$ and several values of the
reinjection parameter $m$, while Fig. \ref{fig:3}(b) shows the same plot with
a fixed value of $m$ and for different values of the nonlinearity
$z$. Thus, relation Eq.~(\ref{scale}) is satisfied independently for
each parameter $z$ and $m$. The critical exponent $\beta$ as a
function of $z$ and as a function of $m$ can be calculated from the
slopes of each curve in Fig. \ref{fig:3}(a) and \ref{fig:3}(b), respectively.

\begin{figure}[th]
\vspace{0.3cm}
\centerline{
\includegraphics[scale=1.15]{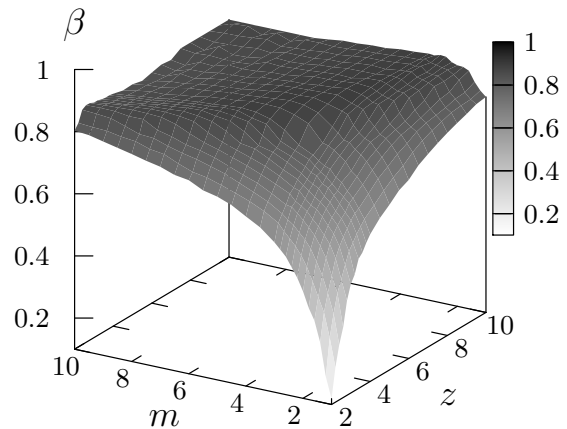}}
\caption{Critical exponent $\beta$ as a function of $z$ and $m$.}
\label{fig:4}
\end{figure}

Figure \ref{fig:4} shows the resulting graph of the critical exponent $\beta$
as a function of both $z$ and $m$. The  values of $\beta$ are
distributed continuously on the interval $(0,1)$. We find the value
$\beta=0.495\pm 0.005$ (typical error) for the parameter values $z=3$ and
$m=1$, for which Eq.~(\ref{map}) corresponds to the tangent map originally
associated to type-III intermittency and having uniform reinjection
probability, as assumed by  Pomeau and Manneville \cite{Pomeau,Manneville}.
However, for other values of $z$ and $m$ there is a manifested deviation from
the Pomeau-Manneville's prediction $\beta=1/2$. The calculated range for the
critical exponent includes the particular value of $\beta \approx 0.79$
obtained for a perturbed logistic map showing type-III intermittency
\cite{Mayer2}.

\section*{3. Conclusions}
We have characterised the scaling behaviour of the
Lyapunov exponent at the transition to chaos in a generic map
exhibiting type-III intermittency by means of a critical exponent
$\beta$. We have shown that $\beta$ depends on both, the order of
the tangency of the map, described by the parameter $z$, and the
probability of reinjection into the laminar region, expressed by the
parameter $m$. We have calculated numerically the critical exponent
on the space of parameters $(z,m)$ and have obtained that $\beta$
varies smoothly on the interval $(0,1)$, in contrast with earlier
predictions for the scaling behaviour of the Lyapunov exponent in
type-III intermittency. It is to be expected that the scaling
behaviour shown here for the Lyapunov exponent could also be detected
in experimental situations with type-III intermittent chaos, where
deviations from the scaling properties of the average laminar length
have already been found.

\section*{Acknowledgements}

This work was supported by CDCHT, Universidad de Los Andes,
M\'erida, and by FONACIT (Venezuela) under grants C-1396-06-05-B and
F-2002000426, respectively. Calculations were performed at CEMVICC,
Universidad de Carabobo, Valencia (Venezuela). M. G. C. thanks the
Consortium of The Americas for Interdisciplinary Science at the
University of New Mexico, Albuquerque, U.S.A., for hospitality while
part of this work was carried out.

\end{document}